\documentclass[twocolumn]{aastex631}

\usepackage{xspace}
\usepackage[nolist]{acronym}
\usepackage{graphicx}
\usepackage{subfigure}
\providecommand{\acrolowercase}[1]{\lowercase{#1}}

\newcommand{\pastro}{$p_\mathrm{astro}$\xspace}

\usepackage{amsmath}
\begin{acronym}
\acro{KDE}[KDE]{kernel density estimation}
\acro{BSN}[BSN]{signal-to-noise coherent Bayes Factor}
\acro{BCI}[BCI]{the coherent versus incoherent Bayes Factor}
\acro{TPR}[TPR]{True Positive Rate}
\acro{FPR}[FPR]{False Positive Rate}
\acro{BNS}[BNS]{binary neutron star}
\acro{NSBH}[NSBH]{neutron star-black hole}
\acro{MDC}[MDC]{Mock Data Challenge}
\acro{EoS}[EoS]{equation of state}
\acro{NS}[NS]{neutron star}
\acro{GW}[GW]{gravitational-wave}
\acro{LIGO}[LIGO]{Laser Interferometer \acs{GW} Observatory}
\acro{aLIGO}[aLIGO]{Advanced \acs{LIGO}}
\acro{AdVirgo}[AdVirgo]{Advanced Virgo}
\acro{O4}[O4]{\ac{aLIGO}'s, \ac{AdVirgo}'s and KAGRA's fourth observing run}
\acro{LVK}[LVK]{LIGO-Virgo-KAGRA}
\acro{IGWN}[IGWN]{International Gravitational-Wave Network}

\acro{O1}[O1]{\ac{aLIGO}'s first observing run}
\acro{O2}[O2]{\ac{aLIGO}'s and \ac{AdVirgo}'s second observing run}
\acro{O3}[O3]{\ac{aLIGO}'s and \ac{AdVirgo}'s third observing run}

\acro{O5}[O5]{\ac{aLIGO}'s and \ac{AdVirgo}'s fifth observing run}
\acro{oLIB}[\acrolowercase{o}LIB]{Omicron+\acl{LIB}}
\acro{KAGRA}[KAGRA]{KAmioka GRAvitational\nobreakdashes-wave observatory}
\acro{GraceDB}[\texttt{GraceDB}]{GRAvitational-wave Candidate Event DataBase}
\acro{LLAI}[LLAI]{Low-Latency Alert Infrastructure}
\acro{CBC}[CBC]{compact binary coalescence}
\acro{BBH}[BBH]{binary black hole}
\acro{FAR}[FAR]{false alarm rate}
\acro{GstLAL}[GstLAL]{GStreamer LIGO Scientific Collaboration Algorithm Library}
\acro{SVD}[SVD]{singular value decomposition}
\acro{SNR}[SNR]{signal\nobreakdashes-to\nobreakdashes-noise ratio}
\acro{FGMC}[FGMC]{Farr, Gair, Mandel and Cutler formalism of Poisson Mixture Model}
\acro{GRB}[GRB]{Gamma Ray Burst}
\acro{PyCBC}[PyCBC]{Python based Compact Binary Coalescence search software}
\acro{MBTA}[MBTA]{Multi-Band Template Analysis}
\acro{SPIIR}[SPIIR]{Summed Parallel Infinite Impulse Response}
\acro{GPU}[GPU]{Graphical Processing Units}
\acro{cWB}[\acrolowercase{c}WB]{Coherent WaveBurst}
\acro{GCN}[GCN]{General Coordinates Network}
\acro{SCiMMA}[SCiMMA ]{Scalable CyberInfrastructure for Multi-Messenger Astrophysics}
\acro{ML}[ML]{Machine Learning}

\acro{MMA}[MMA]{Multi-Messenger Astronomy}
\acro{BAYESTAR}[\texttt{BAYESTAR}]{BAYESian TriAngulation and Rapid localization}
\acro{FITS}[FITS]{Flexible Image Transport System}
\acro{HEALPix}[HEALP\acrolowercase{ix}]{Hierarchical Equal Area isoLatitude Pixelization}
\acro{MOC}[MOC]{Multi-Order Coverage}
\acro{EM}[EM]{electromagnetic}
\acro{NS}[NS]{neutron star}
\acro{BH}[BH]{black hole}
\acro{LLOID}[LLOID]{Low Latency Online Inspiral Detection}
\acro{IIR}[IIR]{infinite impulse response}
\acro{ToO}[ToO]{target of opportunity}
\acro{OPA}[OPA]{Open Public Alert}
\acro{PE}[PE]{Parameter Estimation}
\acro{Bilby}[Bilby]{Bayesian Inference Library for \acs{CBC} \acs{GW} signal}
\acro{RAVEN}[RAVEN]{Rapid, on-source VOEvent Coincident Monitor}
\acro{KNN}[KNN]{K-Nearest Neighbors}
\acro{NN}[NN]{Neural Network}
\acro{MLP}[MLP]{Multilayer Perceptron}
\acro{RF}[RF]{Random Forest}
\acro{GWTC}[GWTC]{Gravitational-Wave Transient Catalog}
\acro{ROC}[ROC]{Receiver Operating Characteristic}
\acro{AUC}[AUC]{Area Under the Receiver Operating Characteristic Curve}
\acro{PDF}[PDF]{probability density function}
\end{acronym}

\begin{document}

\title{Astrophysical or Terrestrial: Machine learning classification of gravitational-wave candidates using multiple-search information}

\author[0009-0000-0499-8988]{Seiya Tsukamoto}
\affiliation{School of Physics and Astronomy, University of Minnesota, Minneapolis, Minnesota 55455, USA}
\author[0009-0008-9546-2035]{Andrew Toivonen}
\affiliation{School of Physics and Astronomy, University of Minnesota, Minneapolis, Minnesota 55455, USA}
\author[0009-0001-8352-8008]{Holton Griffin}
\affiliation{School of Physics and Astronomy, University of Minnesota, Minneapolis, Minnesota 55455, USA}
\author[0009-0005-9422-0091]{Avyukt Raghuvanshi}
\affiliation{School of Physics and Astronomy, University of Minnesota, Minneapolis, Minnesota 55455, USA}
\author[0009-0003-3658-9080]{Megan Averill}
\affiliation{School of Physics and Astronomy, University of Minnesota, Minneapolis, Minnesota 55455, USA}
\author[0009-0000-4242-7406]{Frank Kerkow}
\affiliation{School of Physics and Astronomy, University of Minnesota, Minneapolis, Minnesota 55455, USA}
\author[0000-0002-8262-2924]{Michael W. Coughlin}
\affiliation{School of Physics and Astronomy, University of Minnesota, Minneapolis, Minnesota 55455, USA}
\author[0009-0009-6826-4559]{Man Leong Chan}
\affiliation{Department of Physics and Astronomy, The University of British Columbia, Vancouver, BC V6T 1Z4, Canada}
\author[0000-0001-9898-5597]{Leo Singer}
\affiliation{Astroparticle Physics Laboratory, NASA Goddard Space Flight Center, Greenbelt, MD 20771, USA}
\affiliation{Joint Space-Science Institute, University of Maryland, College Park, MD 20742, USA}


\newcommand{\Bayestar}{\texttt{BAYESTAR}}

\newcommand{\msun}{\ensuremath{M_{\odot}}}
\newcommand{\mc}{\ensuremath{\mathcal{M}_c}}
\newcommand{\hubbleunit}{\ensuremath{\text{km s}^{-1}~\text{Mpc}^{-1}}}
\newcommand{\totalinjections}{\ensuremath{5\times 10^4}}
\newcommand{\BNSrec}{\ensuremath{1489}}
\newcommand{\NSBHrec}{\ensuremath{1105}}
\newcommand{\BBHrec}{\ensuremath{1920}}

\newcommand{\BNSinj}{\ensuremath{40.9\%}}
\newcommand{\NSBHinj}{\ensuremath{35.8\%}}
\newcommand{\BBHinj}{\ensuremath{23.3\%}}

\newcommand{\hasns}{\ensuremath{\texttt{HasNS}}}
\newcommand{\hasremnant}{\ensuremath{\texttt{HasRemnant}}}
\newcommand{\hasmassgap}{\ensuremath{\texttt{HasMassGap}}}

\newcommand{\tsuperevent}{t_{\text{superevent}}}
\newcommand{\tmerger}{t_{\text{0}}}
\newcommand{\tevent}{t_{\text{event}}}
\newcommand{\tadvreq}{t_{\text{ADV\_REQ}}}
\newcommand{\tgcn}{t_{\text{GCN\_PRELIM}}}
\newcommand{\temcoinc}{t_{\text{EM\_COINC}}}
\newcommand{\travenalert}{t_{\text{RAVEN\_ALERT}}}

\newcommand{\publicalertthreshold}{\ensuremath{ \leq 2.31 \times 10^{-5}}\,Hz (two per day)}
\newcommand{\significantalertthreshold}{\ensuremath{\leq 3.9 \times 10^{-7}}\,Hz (one per month)}
\newcommand{\significantindividualalertthreshold}{\ensuremath{\leq 6.4 \times 10^{-8}}\,Hz (one per 6 months)}
\newcommand{\significantindividualalertthresholdcurrent}{\ensuremath{\leq 7.7 \times 10^{-8}}\,Hz (one per 5 months)}
\newcommand{\burstalertthreshold}{\ensuremath{\leq 3.2 \times 10^{-8}}\,Hz (one per year)}

\newcommand{\dc}[1]{{\textcolor{red}{{[DC: {#1}]}} }}
.

\begin{abstract}
{
Low-latency gravitational-wave alerts provide the greater multi-messenger community with information about the candidate events detected by the \ac{IGWN}. Prompt release of data products such as the sky localization, \ac{FAR}, and \pastro values allow astronomers to make informed decisions on which candidate gravitational-wave events merit \ac{ToO} follow-up. However, false alarms, often referred to as ``glitches", where a gravitational-wave candidate, or trigger, is the result of terrestrial noise, are an inherent part of gravitational-wave searches. In addition, with the presence of multiple gravitational-wave searches, different searches may have varying assessments of the significance of a given trigger. As a complement to quantities such as \pastro, we provide a \ac{ML} based approach to determining whether candidate events are astrophysical or terrestrial in nature, specifically a classifier that utilizes information provided by multiple low-latency search pipelines in its feature space. This classifier has a performance an \ac{AUC} of 0.96 and accuracy of 0.90 on the Mock Data Challenge training set and an \ac{AUC} of 0.93 and accuracy of 0.86 on events from the \ac{O3}.}
\end{abstract}



\section{Introduction} \label{sec:intro}

The first direct detection of gravitational waves, which originated from the \ac{BBH} merger event GW150914 \citep{LIGOScientific:2016aoc}, provided a new means for studying the universe. The subsequent joint detection of gravitational waves from the \ac{BNS} merger GW170817 \citep{AbEA2017b} and the optical counterparts in kilonova AT2017gfo \citep{CoFo2017,SmCh2017,AbEA2017f} and GRB170817A \cite{LIGOScientific:2017zic, Goldstein:2017mmi, Savchenko:2017ffs} led to enormous growth of interest in multi-messenger astronomy and \ac{GW} follow-up searches. Observations of kilonovae, astrophysical transients that can be produced by \ac{BNS} or \ac{NSBH} mergers, are of great interest across nuclear astrophysics \citep{Margutti:2017cjl, Smartt:2017fuw, KaNa2017, Kasen:2017sxr, Watson:2019xjv}, cosmology \citep{Abbott:2017xzu, Coughlin:2019vtv, Dietrich_2020}, and for tests of General Relativity \citep{Ezquiaga:2017ekz, Baker:2017hug, Creminelli:2017sry}. Observations from GW170817 and AT2017gfo specifically revealed r-process nucleosynthesis took place and powered kilonova emissions, as evidenced by the presence of r-process elements in the ejecta post-merger \citep{LaSc1974,LiPa1998,MeMa2010,KaMe2017}.



In this era of multi-messenger astronomy, where we hope to combine \ac{GW} detections with \ac{EM} or neutrino observations, the timely release of \ac{GW} alerts and derived data products is essential. The search for \ac{GW} events and their associated counterparts \citep{KAGRA:2013rdx} continues with \ac{O4}, which began on May 23,  2023\footnote{\url{https://observing.docs.ligo.org/plan}} and is in progress as of the time of writing. Kilonovae are faint and short lived, and can have large sky localizations \citep{Rover2007a, Fair2009,Fair2011,Grover:2013,WeCh2010,SiAy2014,SiPr2014,BeMa2015,EsVi2015,CoLi2015,KlVe2016}, making them difficult targets for follow-up. With these challenges in mind, sending out accurate, low-latency alerts is important for maximizing our chances of locating a transient and observing the peak of emissions.

Prompt release of \ac{GW} candidate event information relies on \ac{GW} search pipelines and the \ac{LVK}'s \ac{LLAI} \citep{LIGOScientific:2019gag, Chaudhary:2023vec}. Alerts are distributed both via GCN\footnote{\url{https://gcn.nasa.gov/}} and the Scalable Cyberinfrastructure to support Multi-Messenger Astrophysics\footnote{\url{https://scimma.org/}} (SCiMMA) and come in two types: notices, which are machine-readable and come in a variety of formats, and GCN circulars, which are human-readable. 

There are two categories of low-latency \ac{GW} searches: ``modeled'' \ac{CBC} \citep{KAGRA:2021vkt} searches and ``unmodeled'' Burst \citep{KAGRA:2021tnv} searches. Modeled \ac{CBC} searches target waveforms from \ac{BNS}, \ac{NSBH}, or \ac{BBH} mergers, while unmodeled searches look for \ac{GW} signals of generic morphologies from a wide variety of astrophysical sources, which may include core-collapse of massive stars, magnetar star-quakes, and other sources, in addition to \ac{CBC}s \citep{KAGRA:2021tnv,KAGRA:2021bhs}. Candidate \ac{GW} events that meet a certain FAR threshold are released publicly, in the form of \textit{low-significance} and \textit{significant} alerts \citep{Chaudhary:2023vec}. Up-to-date information on the current \ac{FAR} threshold and the searches running in low-latency can be found in the \ac{LVK} userguide\footnote{\url{https://emfollow.docs.ligo.org/userguide/}}. 

Significant alerts, those which pass the most stringent FAR threshold, are often the subject of downstream analyses and targets of \ac{ToO} searches. Candidate events may be uploaded by multiple search pipelines. When these events are temporally coincident around the same signal, they are grouped together into what is termed a \textit{superevent}. This superevent corresponds to a single candidate \ac{GW} detection, and the individual candidate event with the highest \ac{SNR} that passes the \ac{FAR} threshold becomes the \textit{preferred event}. When passing the public \ac{FAR} threshold, certain information about preferred event becomes public, and is the subject of our alerts. It is possible for the preferred event to change as new events are uploaded, and given the different search pipelines, it is also possible that different events within a superevent are not in perfect agreement on certain aspects of the event.

If after additional analyses a significant candidate event is deemed to not be of astrophysical origin, a retraction is issued for that candidate event. As opposed to astrophysical events, these events are referred to as ``terrestrial", meaning they have resulted from sort of environmental or detector noise source. Terrestrial events, often referred to as noise triggers, and their eventual retraction are an inherent part of \ac{GW} searches as setting a FAR threshold defines your tolerance for false alarms and the rate at which you expect them. However, minimizing the number of retractions, promptly retracting noise triggers, and providing information on the likelihood a candidate event is astrophysical are all important parts of improving the reliability of \ac{GW} candidates.

There have been over 200 significant \ac{GW} candidate events detected during \ac{O4}, excluding retractions.  While only a small fraction of significant events are retracted--alongside the first 200 significant events were 25 retractions--there is demand for data products for assessing confidence in \ac{GW} candidate events. Certain events, such as those likely to include at least one \ac{NS}, are often targets of extensive \ac{ToO} follow-up, using notable person-power and telescope time. Avoiding using such resources on noise triggers is important for efficient follow-up of 
\ac{GW} events, especially looking forward to next generation \ac{GW} detectors such as LISA \citep{Baker:2019nia}, Cosmic Explorer \citep{Reitze:2019iox, Evans:2021gyd}, and  and Einstein Telescope \citep{Abac:2025saz} where the expected rate of events is much higher.

For training our classifiers, we focus on four \ac{CBC} searches which participated in the pre-\ac{O4} \ac{MDC} \citep{Chaudhary:2023vec}: (\ac{GstLAL} ~\citep{Messick:2016aqy, Tsukada:2023edh, Ewing:2023qqe}, \ac{MBTA} \citep{Aubin:2020goo}, \ac{PyCBC} \citep{Nitz:2018rgo,DalCanton:2020vpm}, and \ac{SPIIR} ~\citep{Chu:2020pjv, Kovalam:2021bgg}). Each search pipeline has independent methods of estimating the noise background and calculating quantities such as \acs{FAR} and \pastro, the probability of astrophysical origin; for details see \cite{Messick:2016aqy, Aubin:2020goo, Hooper:2011rb,Luan:2011qx, Usman:2015kfa, DalCanton:2020vpm}. There have also been efforts to combine \pastro information from multiple search pipelines for a ``unified'' \pastro \citep{Banagiri:2023ztt}, which would provide a similar use as our classifier.

Alongside the search pipelines, the \ac{LLAI} encompasses parameter estimation and source classification efforts which are included in alerts and crucial for multi-messenger searches \citep{SiPr2016, Ashton:2018jfp, Chatterjee:2019avs, Berbel:2023vug}. Additional machine learning-based searches \cite{Marx:2024wjt} and rapid parameter estimation techniques are also in development \citep{Dax:2021tsq, Chatterjee:2024pbj}. Currently, pipeline \pastro values and \texttt{GWSkyNet} \citep{Cabero:2020eik, 2022ApJ...927..232A, Raza2023, Chan:2024kzu, Raza:2025ouw} provide statements of confidence for the astrophysical nature of candidate events; we envision our classifier(s) as a complementary data product to those efforts, in addition to possible future data products aimed at multi-messenger sources \citep{Berbel:2023vug, Toivonen:2024ike}.

\section{Data}

\subsection{Mock Data Challenge}

The \ac{MDC} \citep{Chaudhary:2023vec} is a real-time simulation campaign which was carried out in anticipation of \ac{O4} to stress-test the \ac{LLAI} and evaluate the performance of \ac{GW} searches and data products. The \ac{MDC} takes a 40 day stretch of background data taken by \ac{LVK} interferometers during \ac{O3}, and injects \ac{CBC} waveforms across the \ac{BNS}, \ac{NSBH}, and \ac{BBH} parameter space.  A total of {\totalinjections} simulated \ac{CBC} waveforms of varying mass and spins were injected into the aforementioned stretch \ac{O3} of data, with the \ac{GW} search pipelines running on the data stream. This results in thousands of recovered injections with significant \ac{FAR}s, and a realistic dataset for downstream analyses. 

Due to the numerous recovered injections and comprehensive nature of the dataset, we use this \ac{MDC} as our training data. However, we must note the that the rate of injections is artificially high and not representative of the true discovery rate in \ac{O4}. This impacts \ac{FAR} values, and consequentially \pastro calculations reliant on those \ac{FAR} values. As certain searches rely on a local background estimation for \ac{FAR} calculations, their \ac{FAR} values are subject to a upward bias due to large density of high \ac{SNR} injections in the data \citep{Chaudhary:2023vec}. As discussed later in Sec. \ref{sec:O3}, this will impact the performance of our classifier when evaluating accuracy on real candidate events from an observing run, such as those from \ac{O3}. Resolving this upward bias in \ac{FAR} is a high priority issue ahead of the \ac{MDC} that will be carried out in anticipation of \ac{O5}, using a stretch of \ac{O4} data. One proposed solution would be to have pipelines evaluate the background based on the data stream alone, without any injections present. 
\\ \\



\subsection{\ac{O3} GW Detections}

We also evaluate the performance of the classifier on real \ac{GW} events from a past observing run, to ensure the model can still make reliable predictions for these real events. In order to assess our accuracy, however, we must assume true labels for our dataset, and provide events with true labels of both astrophysical and terrestrial. We assign all pipeline events corresponding to a confirmed significant \ac{O3} event in the \ac{GWTC} catalog \citep{KAGRA:2021vkt} a true label of astrophysical, and take the retracted significant events as our terrestrial class. This is a reasonable assumption, as all events in the \ac{GWTC} catalog have undergone significant human-vetting, while retracted events can be assumed to be terrestrial due to their data quality issues, glitches present in the strain, and lack of significant \ac{GW} signal. As a note for \ac{O3} events, since a superevent is a grouping of individual pipeline events, there are some individual pipeline events which correspond to a significant superevent that do not pass the significant \ac{FAR} threshold on their own. We still assign these events a true label of astrophysical and evaluate our model on them. In the future, once candidate events from \ac{O4} have been added to the \ac{GWTC} catalog, we can also release the performance on those events.

\section{The Bayes Factors}

We take a Bayesian approach. Bayes theorem is stated as:
\begin{equation}
    P(H_i|D) = \frac{P(D|H_i) \cdot P(H_i)}{P(D)},
\end{equation}

where $D$ is the \ac{GW} event and $H_i$ is the hypothesis. $P(H_i|D)$ and $P(D|H_i)$ are conditional probabilities defined as the posterior and likelihood under hypothesis $i$. $P(H_i)$ and $P(D)$ are defined as the prior and marginal probabilities. As the marginal is constant throughout different hypotheses, the relation between the ratio of priors and posteriors between hypothesis' can be described through the Bayes Factor: 

\begin{equation}
    BF_{i,j} =  \frac{P(D|H_i)}{P(D|H_j)}.
\end{equation}

\texttt{BAYESTAR} provides the \ac{BSN} \citep{Veitch:2009hd, SiPr2016} which is defined as:

\begin{equation}
    \log(BSN) =  \log\left( \frac{P(D_{HL}|H_{signal+noise})}{P(D_{HL}|H_{signal})}\right).
\end{equation}
\texttt{BAYESTAR} also provides \ac{BCI} which is defined as:
\begin{equation}
    \log(BCI) = \frac{\log(BSN)}{\sum \log(BSN_{incoherent})},
\end{equation}
where $BSN_{incoherent}$ is the \ac{BSN} of each interferometer independently. These are the two Bayes Factors that will be used to aid in classifying the \ac{GW} events from the \ac{MDC} data \citep{Chaudhary:2023vec}. 

\subsection{FAR and SNR}
\ac{FAR} is a measure of significance produced by each of the pipeline for each potential candidate \citep{AbEA2019}, this metric helps determine whether or not we send out an alert for a candidate event, and lower FAR values are correlated with events of astrophysical nature. This is defined as:
\begin{equation}
    FAR(\rho) = \int_{\rho}^{\rho_{\text{max}}} \Lambda_n p_n (\rho')(d\rho'),
\end{equation}
where $\Lambda_n$ is the mean Poisson rate of signal and noise triggers, $\rho$ is the \ac{SNR}, and $p_n$ is the \ac{PDF} \citep{Callister:2017}. The \ac{SNR}, signal-to-noise ratio, provides an estimate of the signal strength is by comparing it to the noise produced in a given detection. High \ac{SNR} events are correlated with events of astrophysical nature. However, glitches with high power and non-Gaussian distribution can cause large \ac{SNR} to be present in non-astrophysical triggers \citep{PhysRevLett.119.161101}. \ac{SNR} is correlated with the \ac{BSN} as seen in Fig. \ref{fig:corner_gstlal}. The \ac{FAR} and \ac{SNR} will be implemented as additional features for the classification algorithms used.

\subsection{\pastro}
The probability of astrophysical origin, termed \pastro, is provided for \ac{GW} event candidates and used to help determine whether an event is astrophysical or terrestrial in nature. \pastro is dependent on \ac{FAR} and \ac{SNR}, and each search pipeline calculates \pastro differently \citep{Chaudhary:2023vec}. Our data product is meant to be complementary to \pastro, and we later evaluate our model performance with \pastro as a baseline for comparison in Sec. \ref{Results}.



\section{Model Architectures}\label{sec:architectures}
Noise triggers are often single-pipeline events, so there is little overlap between \ac{CBC} pipelines on terrestrial events. Because of this, the relationship between all data products from different \ac{CBC} pipelines was not able to be modeled in a single architecture. In contrast, \texttt{BAYESTAR} is run on every event, so the relationship between its data products, specifically the BSN and BCI, with individual \ac{CBC} search pipelines was able be modeled. In order to resolve this limitation, classification was split into two stages. In the first stage, we classify using each \ac{CBC} search pipeline's preferred event and \texttt{BAYESTAR}'s data products in order to assign scores between 0 and 1 for each pipeline. If a pipeline does not recover an event, a value of $-1$ is assigned to indicate the lack of detection. In the second stage, the individual scores were used to assign a final score for the superevent using the same model architecture for both stages of classification. The hyperparameters are shown in appendix Sec. \ref{sec:hyperparameters}, and were each tuned through cross-validation grid search.  

\subsection{K-Nearest Neighbors}

\ac{KNN} is an algorithm that makes predictions based on the proximity of a test data point in the parameter space to the data points in the training set. The algorithm assigns labels and probabilities based on a set of nearest neighbors of size $K$ defined by a distance metric. We use the open-source Python implementation of \ac{KNN} from \texttt{scikit-learn} for our classifier. More detail is provided in the appendix, Sec. \ref{Appendix:KNN}.

\subsection{Random Forest}

\ac{RF} is an ensemble learning method used for classification by constructing many decision trees during training. The hierarchical structure of the tree starts with a top node that contains the entire data set, with each tree trained independently of each other. At each branch the data is subdivided into two child nodes from a decision that is based on the features given. The splitting repeats, recursively, until it reaches a predefined stopping criteria. The end result is the best prediction given the dataset and the labels. We use the open-source Python implementation of \ac{RF} from \texttt{scikit-learn} for our classifier. Additional details are provided in the appendix, Sec. \ref{Appendix:RF}.

\subsection{Neural Network}

\ac{MLP} is a simple \ac{NN} that contains a set of hidden layers and activation functions that perform linear and non-linear operations on the data. Using the gradient decent optimizer \texttt{ADAM}, neurons in each hidden layer can be trained to reduce binary cross entropy loss. We used the open-source Python implementation of \ac{NN}'s from \texttt{PyTorch} for our classifier. The hyperparameters were tuned through a cross-validated grid search resulting in 10 hidden layers with 59 neurons each for the first stage and 2 hidden layers with 52 neurons each for the second stage. More detail is provided in the appendix, Sec. \ref{Appendix:NN}.


\begin{figure*}[htp]
    \centering
    \includegraphics[width=\textwidth]{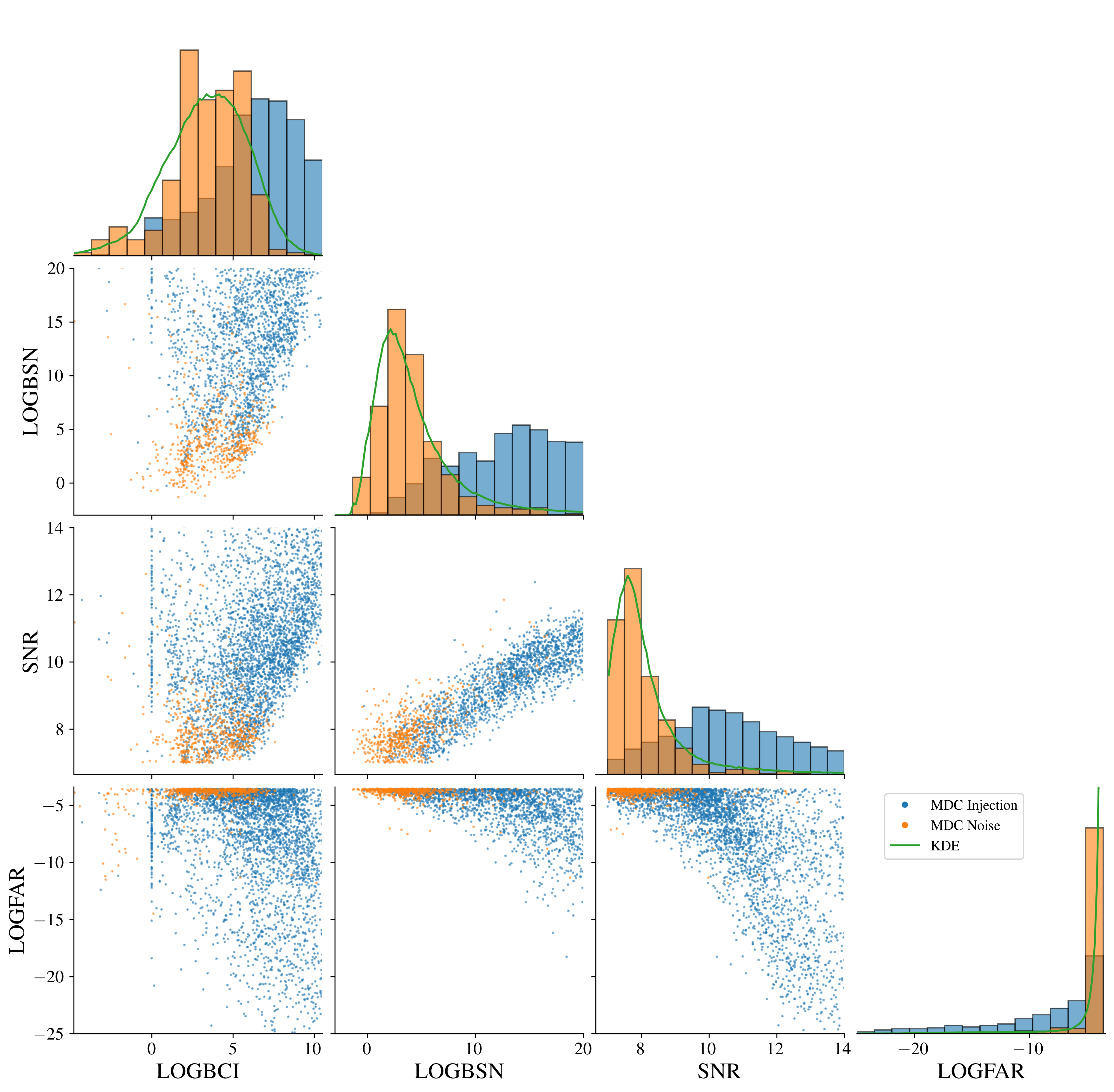}
    \caption{Corner plot of the \ac{GstLAL} and \texttt{BAYESTAR} pipeline data products from \ac{MDC} events windowed around noise events. The histograms are normalized and the green line shows the \ac{PDF} of the KDE.}
    \label{fig:corner_gstlal}
\end{figure*}

\subsection{Feature Selection} \label{sec:features}

We utilize event information from four \ac{CBC} pipelines, \ac{GstLAL}, \ac{MBTA}, \ac{PyCBC}, and \ac{SPIIR}, in our feature space. This multiple pipeline approach allows us to capture the information provided by each of these searches. In addition, events detected by a only single pipeline are much more likely to be terrestrial than those found by multiple pipelines, while events detected by multiple-pipelines are much more likely to be astrophysical. Specifically, we used the highest \ac{SNR} event upload that passes the \ac{FAR} threshold for each pipeline in a superevent. 

We use the following features:\\
\\
The Bayes factors, \ac{BCI} and \ac{BSN}, are derived from the sky map FITS file for each event generated by \texttt{BAYESTAR} \citep{SiPr2016}:
\begin{enumerate}
    \item Logarithm of the Bayes factor for the signal versus noise hypothesis (LOGBSN)
    \item Logarithm of the Bayes factor for the coherence versus incoherence hypothesis (LOGBCI)
\end{enumerate}
And inputs derived from low-latency pipelines: 
\begin{enumerate}
    \item[3.] Preferred SNR
    \item[4.] Logarithm of preferred FAR (LOGFAR)
\end{enumerate}

\subsection{Dataset}

Our dataset is composed of simulated \ac{MDC} events run through \ac{CBC} search pipelines and \texttt{BAYESTAR} \citep{Chaudhary:2023vec}. The events are then labeled astrophysical or terrestrial by cross-matching the detected events with the set of injections. \ac{FAR} and \ac{SNR} were taken from the highest \ac{SNR} event that passes the \ac{FAR} threshold for each pipeline, also referred to as the preferred event. 
The total number of events labeled as astrophysical is 5726 and the total terrestrial is 1834, resulting in a class imbalance. In order to pre-process LOGBSN, LOGBCI, \ac{SNR}, and LOGFAR into a Gaussian distribution, a logit transform was used on LOGFAR, as it non-Gaussian and instead asymptotic towards high values, and a Yeo-Johnson transform was performed on LOGBSN and SNR, as they were a skewed Gaussian distribution. Using the transformed data products, a multivariate Gaussian \ac{KDE} was used to sample an adequate number of terrestrial events. 

A \ac{KDE} is a way to estimate the \ac{PDF} of a variable based on applying kernels at data points. The \ac{KDE} is described by
\begin{equation}
    f(x) = \frac{1}{nh}\sum_{i=1}^{n} K (\frac{x - x_i}{h})
\end{equation}
where $h$ is the kernel bandwidth, and $n$ is the number of data points. We chose a multivariate gaussian kernel, and calibrated the \ac{KDE} for each pipeline separately. The kernel bandwidth was chosen using a leave one out maximum likelihood estimate.

A multivariate \ac{KDE} was created for each pipeline, and an example of one can be seen in Fig. \ref{fig:corner_gstlal} for the pipeline \ac{GstLAL}. With a sufficiently large dataset, we could instead sample events for each combination of the four pipelines, but since 1778 of 1834 terrestrial events are only recovered by a single pipeline, there are too few events for this approach. 
This combination of many more astrophysical events than terrestrial, and the fact that these are not evenly distributed across the different pipeline configurations means there is a large class imbalance. In order to resolve the class imbalance and ensure sufficient astrophysical and terrestrial events for each pipeline configuration, class weights equivalent to the inverse of class size were used in \ac{KNN} and \ac{RF}. Downsampling was used for \ac{NN} while resampling the training set when the model performance plateaus.


\section{Results}\label{Results}


\subsection{Performance of Classifiers on simulated \ac{MDC} Events}\label{sec:MDC}
\begin{figure}[h]
    \centering
    \includegraphics[width=\linewidth]{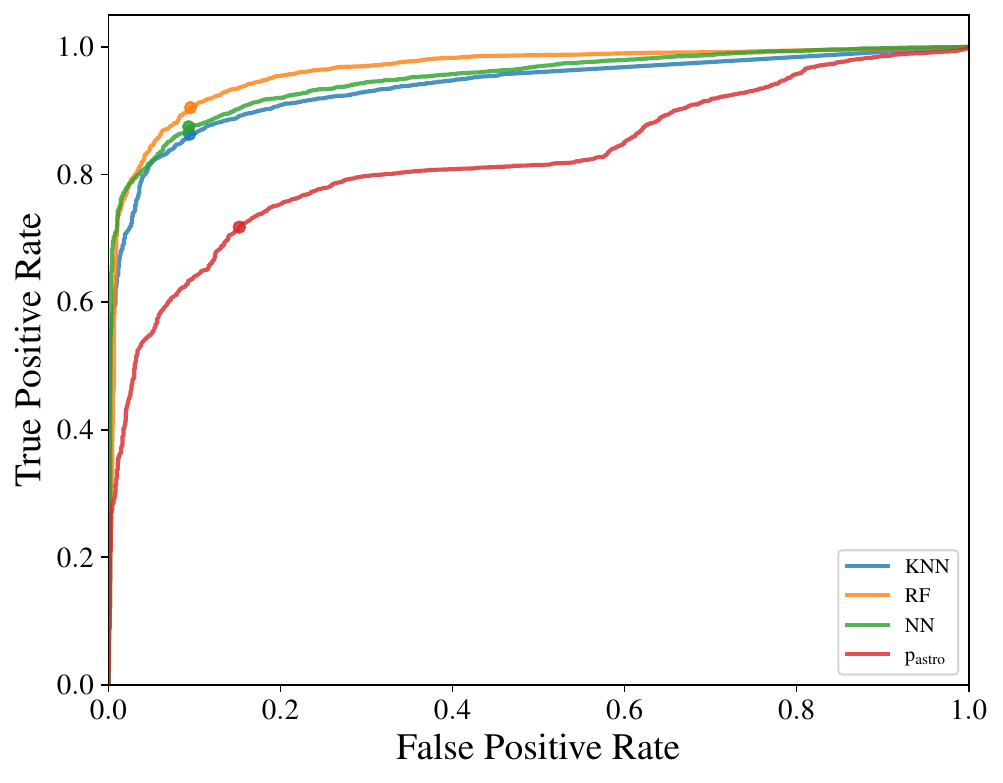}
    \caption{\ac{ROC} curves obtained from the \ac{MDC}. We use the the \pastro from the preferred event of each superevent for comparison. The points on the curve denote the threshold that minimizes the distance to the upper left corner of the plot.}
    \label{fig:roc_mdc}
\end{figure}

\begin{figure}[h]
    \centering
    \includegraphics[width=\linewidth]{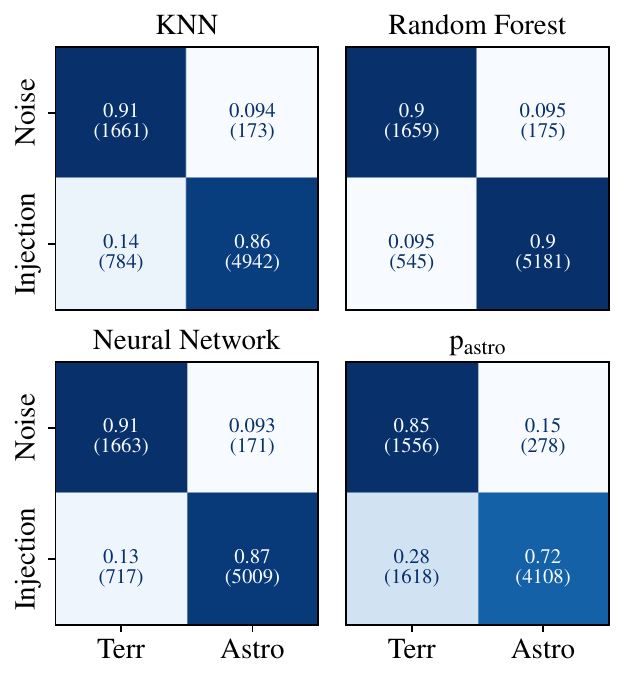}
    \caption{Confusion matrix of model performance on the \ac{MDC} dataset. We used the the \pastro from the preferred event of each superevent for comparison.}
    \label{fig:confusion_mdc}
\end{figure}

Fig.~\ref{fig:roc_mdc} shows the \ac{ROC} curves for the all classifiers using ten fold cross validation. \ac{RF} and \ac{NN} had an \ac{AUC} of 0.96 and 0.95, respectively, outperforming \ac{KNN} which had an \ac{AUC} of 0.94. Each of our classifiers outperform \pastro. The corresponding confusion matrices in Fig. \ref{fig:confusion_mdc} show each classifiers' performance compared to \pastro using a threshold of 0.10, 0.52, 0.37, and 0.50 for \ac{KNN}, \ac{RF}, \ac{NN}, and \pastro respectively. The thresholds of the three classifiers trained here were decided by finding the threshold in the \ac{ROC} that minimized the distance to the top left corner. Since \pastro had large overclassification in opposite classes between \ac{MDC} and \ac{O3}, so the threshold was kept at 0.50. The \ac{RF} and \ac{NN} slightly outperform the \ac{KNN} for astrophysical events, with \ac{RF} outperforming the others for terrestrial events. 

\begin{figure}
    \centering
    \includegraphics[width=1\linewidth]{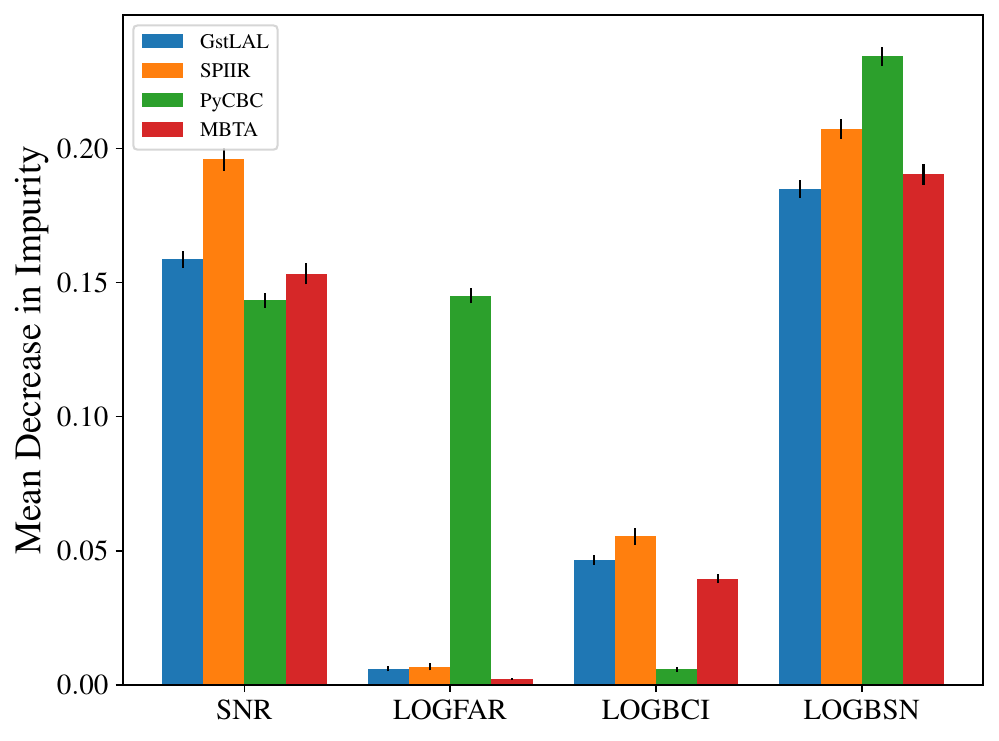}
    \caption{Permutation importance for the \ac{RF} classifier for each pipeline.}
    \label{fig:feat.importance}
\end{figure}

Fig. \ref{fig:feat.importance} shows permutation feature importance of the \ac{RF} model as an example demonstrating that each pipeline has differing feature importance. This validates our approach to have separate models for each pipeline. In general, LOGBSN was found to have the highest feature importance across pipelines. Although permutation tests show that LOGFAR contributes the least to per-pipeline importance, it still provides a marginal gain, particularly for single-pipeline events, so we retain it for completeness. As explained in \cite{Chaudhary:2023vec}, the high rate of injections in the \ac{MDC} leads to biased \ac{FAR} values in our training set, something we hope would be resolved for a future pre-O5 \ac{MDC} and should increase the LOGFAR feature importance.

\begin{figure}[h]
    \centering
    \includegraphics[width=\linewidth]{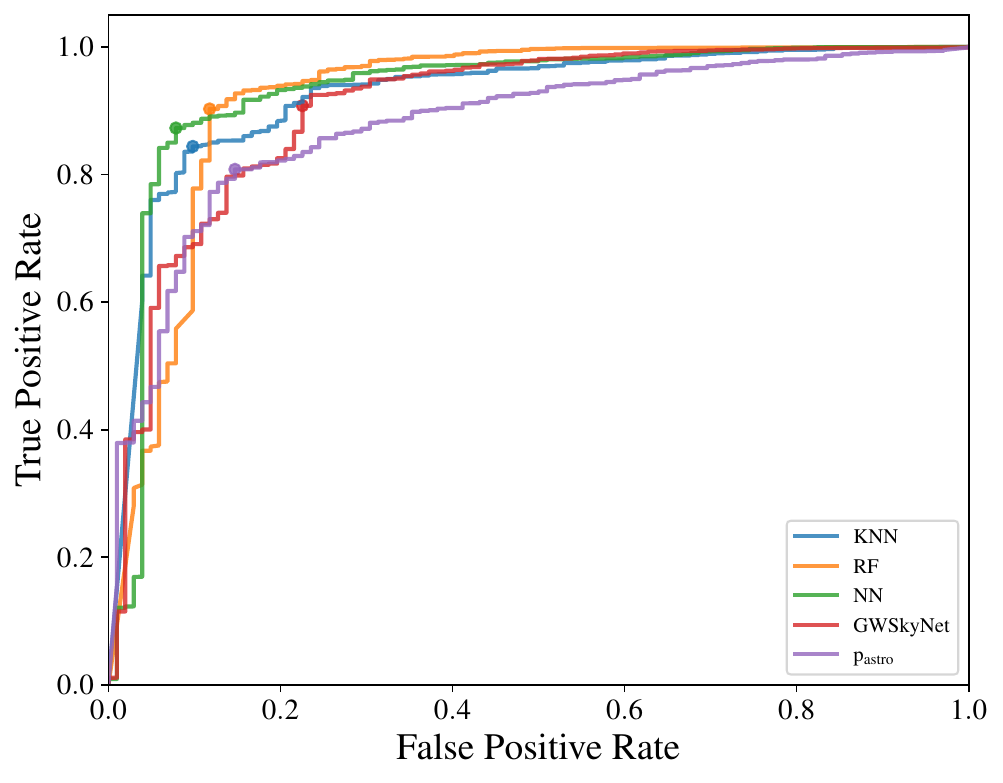}
    \caption{\ac{ROC} curves from the subset of \ac{MDC} events \texttt{GWSkyNet} was run on, specifically those with \ac{FAR} $\leq 2$ per day, network \ac{SNR} $\geq 7$, and two or more individual detector \ac{SNR} $\geq 4.5$. We used the the \pastro from the preferred event of each superevent for comparison. The points on the curve denote the threshold that minimizes the distance to the upper left corner of the plot for this subset of events.}
    \label{fig:roc_GWSN}
\end{figure}

In Fig. \ref{fig:roc_GWSN} we compare our classifier's performance to that of \texttt{GWSkyNet} \citep{Chan:2024kzu}. This plot displays ROC curves for the subset of MDC events \texttt{GWSkyNet} was run on, specifically those with \ac{FAR} $\leq 2$ per day, network \ac{SNR} $\geq 7$, and two or more individual detector \ac{SNR} $\geq 4.5$. We see that our classifiers and \texttt{GWSkyNet} significantly outperform \pastro, and have similar performance to one another on this subset of events. We intend for our classifier to be complementary to \texttt{GWSkyNet}, with the major difference being our classifiers take advantage of multiple-pipeline information.

\subsection{Performance on \ac{O3}}\label{sec:O3}
\begin{figure}[h]
    \centering
    \includegraphics[width=\linewidth]{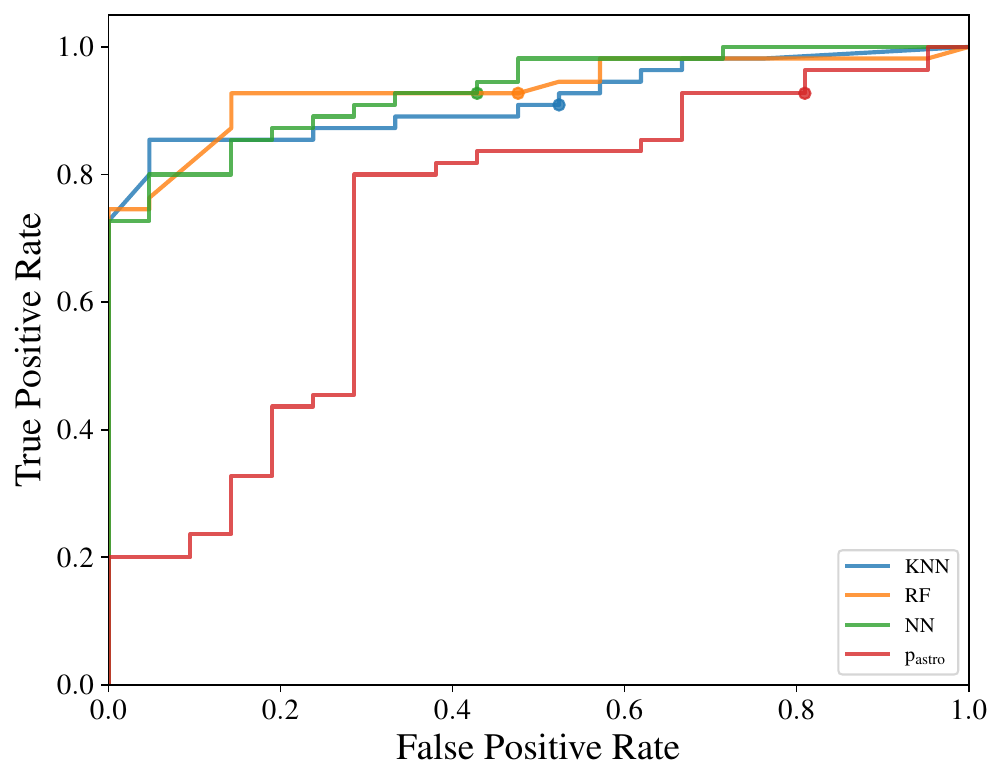}
    \caption{\ac{ROC} curves obtained from the testing on the \ac{O3} dataset. The points on the curve denote the same thresholds described in Sec. \ref{sec:MDC}.}
    \label{fig:roc_O3}
\end{figure}

\begin{figure}[h]
    \centering
    \includegraphics[width=\linewidth]{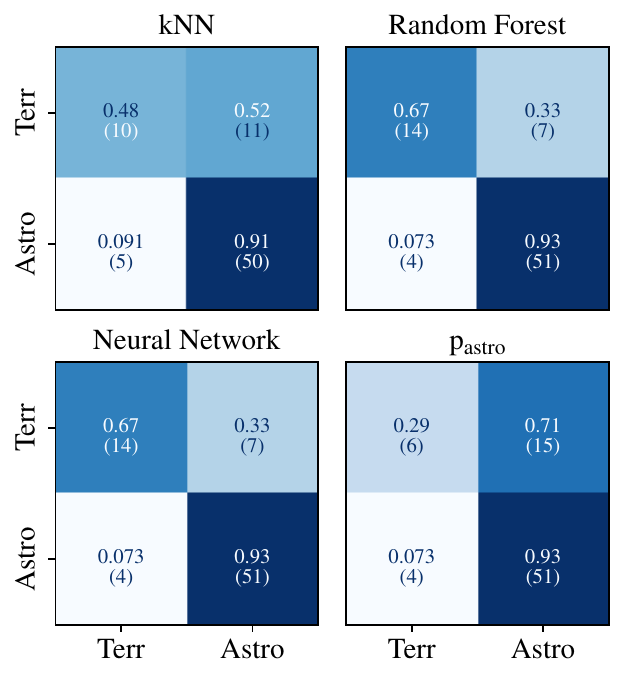}
    \caption{Confusion matrix of model performance on the \ac{O3} dataset.}
    \label{fig:confusion_O3}
\end{figure}

Fig.~\ref{fig:roc_O3} shows the \ac{ROC} curves for the all classifiers on the \ac{O3} data. The \ac{RF} and \ac{NN} have matching performance with an AUC of 0.93 outperforming \ac{KNN} with a performance of 0.92. The corresponding confusion matrix in Fig.~\ref{fig:confusion_O3} shows all classifiers' performance in comparison to \pastro using the same thresholds described in Sec. \ref{sec:MDC}. KNN and RF has disagreement in three astrophysical events and four retractions. RF and NN has disagreement in four astrophysical events and four retractions. NN and KNN has disagreement in three astrophysical events and six retractions.

\begin{figure*}[htp]
    \centering
    \includegraphics[width=\textwidth]{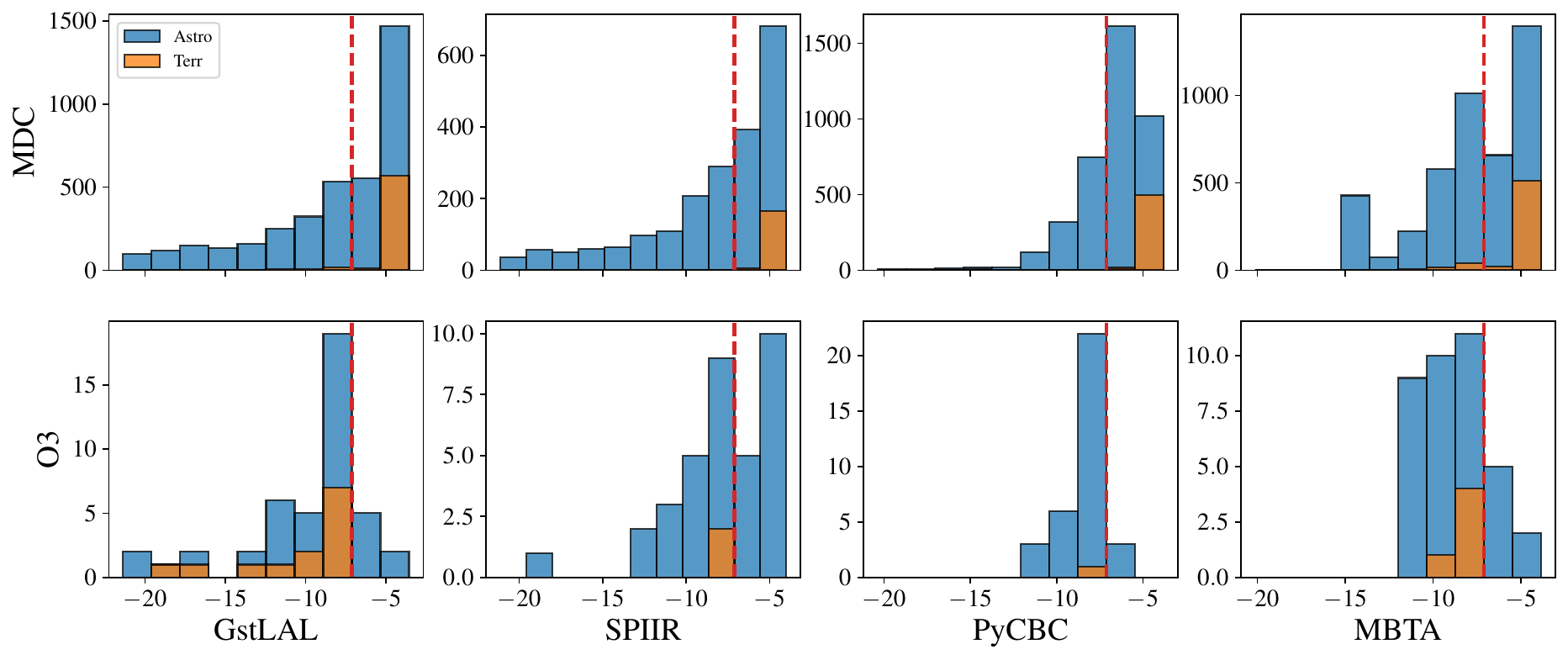}
    \caption{Distributions of \ac{MDC} and \ac{O3} events against LOGFAR and their true labels. The red dotted line shows the five month significant threshold used on O3.}
    \label{fig:far_bias}
\end{figure*}

The true labels for the \ac{O3} dataset are defined differently than for the \ac{MDC} dataset. Confirmed significant events that make the \ac{GWTC} catalog \citep{KAGRA:2021vkt} are given true astrophysical labels, while retracted events are given terrestrial labels. Seeing as each of these events have undergone extensive human-vetting to be a part of the catalog, or to be retracted, this is a fair assumption. However, it is possible there are terrestrial events that have made it into the catalog, which may slightly impact our results. Luckily, training is not impacted as that is done with \ac{MDC} injections. With this in mind, all \ac{O3} superevents must have at least one event with \ac{FAR} greater than the significance threshold, and all individual pipeline's preferred event within that superevent are then considered. There is one exception in \ac{O3}, S200105ae, which was originally subthreshold but published after manual review. All \ac{O3} retractions were single pipeline events meaning there is no subthreshold event labeled as a retraction. \ac{MDC} noise triggers above the significance threshold, corresponding to a LOGFAR of -7.1, are rare, as can be seen in Fig.~\ref{fig:corner_gstlal}. The disparity in the proportion of labels between \ac{MDC} and \ac{O3} is also accentuated by the upwards bias of FAR from the high rate of injections in the \ac{MDC} \citep{Chaudhary:2023vec}. Overall, this bias may cause the classifier for at least one pipeline to have a higher rate of false positives in the \ac{O3} dataset compared to the \ac{MDC} dataset.

Data quality issues in \ac{O3} events can also impact the performance of our model as the model only takes in the data products produced in low latency. High amplitude  glitches can cause terrestrial events to have parameters similar to astrophysical events causing false positives. For example, S191220af had a glitch in L1 with a LOGBSN of 15.56, S191120aj  had glitches in H1 and L1 with a LOGBSN of 27.58, and S191225aq  had glitches in L1 and V1 with a LOGBSN of 110.47. To our model, these strong feature values mimic astrophysical events, even when there are clear data quality issues upon human inspection. Other events, such as S190829u with LOGBSN of 43.80 and S190822c with LOGBSN of 276.94, had no high strain glitches yet can still produce inaccurate data products due to detector background. Astrophysical events with low signal strength account for false negatives as associated values are similar to events due to noise, such as S191205ah, S190910h, and S190718y for the \ac{NN}.

Fig. \ref{fig:far_bias} shows \ac{FAR} distributions of \ac{MDC} and \ac{O3} events for each pipeline, as well as their true labels. We note that there are some \ac{O3} events that do not pass the significant \ac{FAR} threshold, which are those corresponding to confirmed significant superevent in the \ac{GWTC} catalog. These events are correctly labeled regardless of their high 
\ac{FAR}, due to the multiple-pipeline nature of our model.


\section{Conclusion}
In this paper, we present the performance of our classifiers for assessing significance of low-latency \ac{GW} candidate events trained data from the \ac{MDC} \citep{Chaudhary:2023vec}. We cover three distinct model architectures, \ac{KNN}, \ac{RF}, and \ac{NN}, to assess event data products and classify those events as either astrophysical or terrestrial in origin. The \ac{RF} and \ac{NN} classifiers have matching performance with an \ac{AUC} of 0.93 on \ac{O3} events slightly outperforming the \ac{KNN} algorithm with a performance of 0.92. For \ac{MDC} events, the \ac{RF} and \ac{NN} have an \ac{AUC} of 0.96 and 0.95, while the \ac{KNN} is 0.94. In both cases each of our classifiers significantly outperforms \pastro, and could be a useful complementary data product alongside it.

With the limited information available in low-latency, we can reliably determine whether an event is astrophysical or terrestrial, which can be extremely valuable for \ac{EM} follow-up efforts. The most prominent limitation of the classifier is the disconnect between the data products and the detection data quality. The actual 
\ac{GW} strain data is not included in the decision process for each event classification, so data quality issues such as glitches may inflate the significance of our features and lead to some false positives. For example, loud glitches caused false positives in a handful of \ac{O3} events: S191225aq, S191220af, S191120aj, and S190808ae. To improve on this aspect, including a metric for the model to consider a quantitative measure of the data quality of \ac{GW} strain. Another improvement would be a larger and more comprehensive set of terrestrial events to balance the true classes or allow us to more accurately sample the data with a \ac{KDE}. In particular, multiple-pipeline terrestrial events would be useful, but these are rare.

We should also note, that it is possible for a loud terrestrial glitch to be coincident with an astrophysical event, just as was the case with GW170817 \citep{AbEA2017b}. Our classifier has not been explicitly trained or tested on this class of events, and we do not make assumptions about how our model will perform with such events, as quantities such as \ac{FAR} and \ac{SNR} will certainly be impacted. We can also assume extensive human intervention and vetting will be necessary in this case. 

A similar score is produced by \texttt{GWSkyNet}, which is a pipeline using a series of convolutional neural networks trained on \texttt{BAYESTAR} localization and volume sky maps along with numerical inputs including the Bayes factors used here \citep{Cabero:2020eik, Raza2023}. In contrast, our pipeline uses internal information from \textit{multiple} \ac{CBC} search pipelines directly. This means a different parameter space is used compared to \texttt{GWSkyNet}, and the two can make complementary data products alongside \pastro. 

The events in the \ac{MDC} are created by injecting waveforms onto existing \ac{O3} data at a high rate. This causes an upwards bias of \ac{FAR} values if the pipeline calculates \ac{FAR} using an estimation of the local noise background \citep{Chaudhary:2023vec}. This means that when using real \ac{GW} events from \ac{O3}, our performance may be negatively impacted. Luckily, a pre-O5 \ac{MDC} using \ac{O4} data is planned and will hopefully resolve this issue. Retraining and tuning on this dataset will need to be done, as well as evaluating on \ac{O4} events after their public data release. This pre-O5 \ac{MDC} may also be comprised of a different set of \ac{CBC} search pipelines, which is something we can easily accommodate. Any additions can be implemented by adding classifiers in the first stage for additional \ac{CBC} search pipelines and adding parameters for additional post \ac{CBC} search pipelines. We envision this classifier could be used to determine significance of \ac{O5} candidate events alongside other data products such as \pastro and \texttt{GWSkyNet}. This addresses the issue where search pipelines may have differing assessments of significance for a given candidate event and provides a means of combining information from multiple searches into one cohesive statement of significance. We will advocate for the implementation and public release of our classifier's outputs in the future.

\begin{acknowledgements}
We thank Sushant Sharma-Chaudhary for thorough internal review of this paper. We thanks Jess McIver for help with provided \texttt{GWSkyNet} values.
S.T, A.T, H.G., A.R., M.A., F.K., and M.W.C. acknowledge support from the National Science Foundation with grant numbers PHY-2117997, PHY-2308862 and PHY-2409481.
This material is based upon work supported by NSF's LIGO Laboratory which is a major facility fully funded by the National Science Foundation.
\end{acknowledgements}

\software{Astropy \citep{astropy:2013, astropy:2018, astropy:2022}, Bilby \citep{Ashton:2018jfp}, Matplotlib \citep{Hunter:2007}, NMMA \citep{Pang:2022rzc}}

\clearpage
\bibliography{references}
\bibliographystyle{aasjournal}

\begin{appendix}

\section{Hyperparameters}\label{sec:hyperparameters}

\begin{figure}
    \centering
    \includegraphics[width=.5\linewidth]{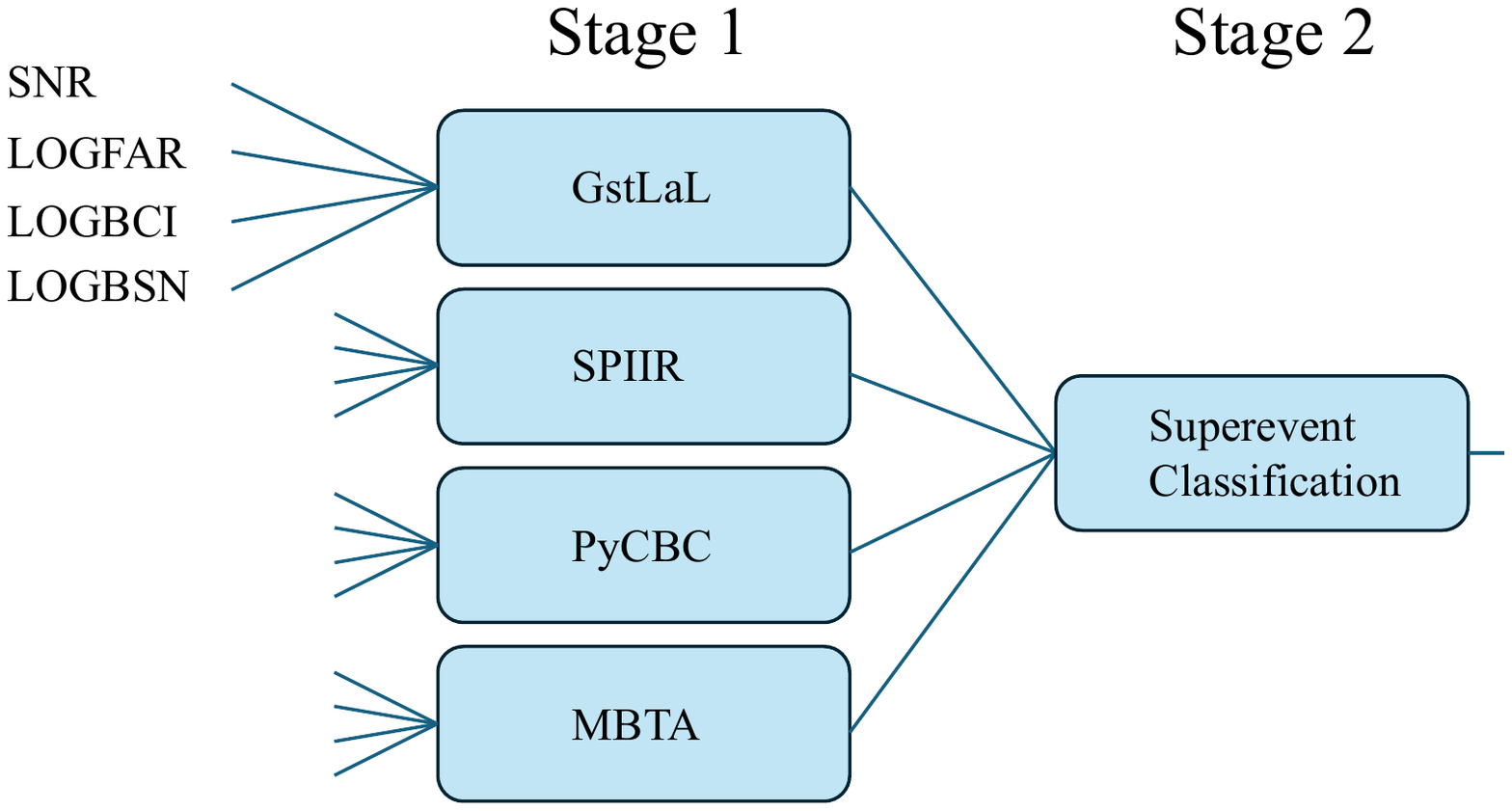}
    \caption{Diagram of model architecture, where each block is either \ac{RF}, \ac{KNN}, or \ac{NN}. The \ac{NN} consists of 10 hidden layers with ReLU activation function in stage 1 and 2 hidden layers with ReLU activation function in stage 2. If a pipeline is not used in a superevent, a value of -1 is sent to superevent classification.}
    \label{fig:Arch}
\end{figure}

Fig. \ref{fig:Arch} outlines the multiple-pipeline model architecture and corresponding features, as covered in Sec. \ref{sec:features}.\\

\subsection{K-Nearest Neighbors (KNN)}\label{Appendix:KNN}

Stage 1 \ac{KNN}: 6 neighbors, distance weight function, and Minkowski metric.

Stage 2 \ac{KNN}: Same as stage 1 except with 693 neighbors.
\\

\subsection{Random Forest (RF)}\label{Appendix:RF}

Stage 1 \ac{RF}: 10 estimators, max depth of 4, Gini criterion, minimum samples to split of 2, minimum samples of each leaf of 1, and maximum features of 4.

Stage 2 \ac{RF}: Same as stage 1 except with 310 estimators and max depth of 7. 
\\

\subsection{Neural Network (NN)}\label{Appendix:NN}

Stage 1 \ac{NN}: 10 hidden layers with 59 neurons each, L2 regularization penalty of 0.0035, learning rate of 0.0063, batch size of 122, ReLU activation function is used for all hidden layers, and binary cross entropy loss. Training is done using a 90/10 train/validation split and training stops when validation loss plateaus.

Stage 2 \ac{NN}: 
2 hidden layers with 52 neurons each, L2 regularization penalty of 1.72e-06, learning rate of 0.008, batch size of 675, ReLU activation function is used for all hidden layers, and binary cross entropy loss. Since train set is redefined at plateau due to downsampling, training stops after the third plateau.

\section{Supporting Figures}

\begin{figure}[b]
     \centering
     \includegraphics[width=.45\linewidth]{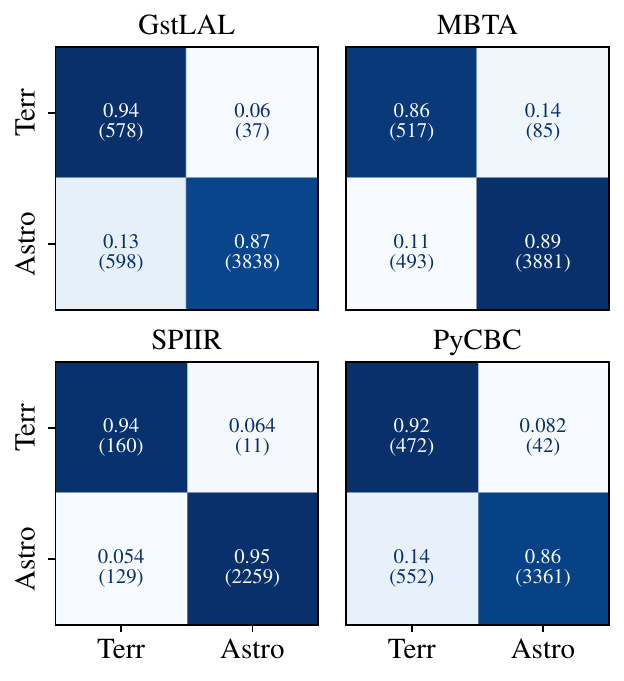}
     \centering
     \includegraphics[width=.45\linewidth]{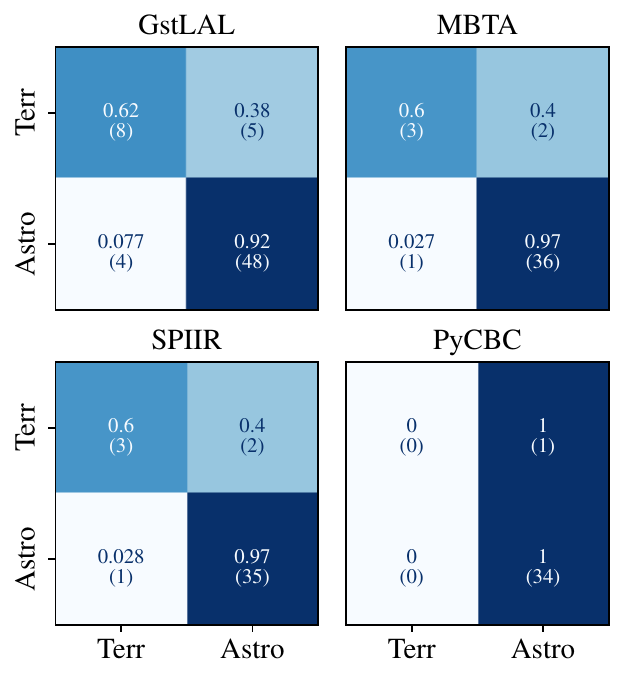}
     \caption{Confusion matrices for \ac{MDC} (left) and \ac{O3} (right) on individual pipelines for \ac{NN}.}     
     \label{fig:cbc_confusion}
\end{figure}

Fig. \ref{fig:cbc_confusion} shows the individual confusion matrices for each of the pipelines trained on for both the \ac{MDC} and \ac{O3}. The individual single-pipeline models have lesser performance on their own, but when combined together as shown in Fig. \ref{fig:Arch} their performance improves significantly, as seen in Figs. \ref{fig:confusion_mdc} and \ref{fig:confusion_O3}.

\begin{figure}
    \centering
    \includegraphics[width=\linewidth]{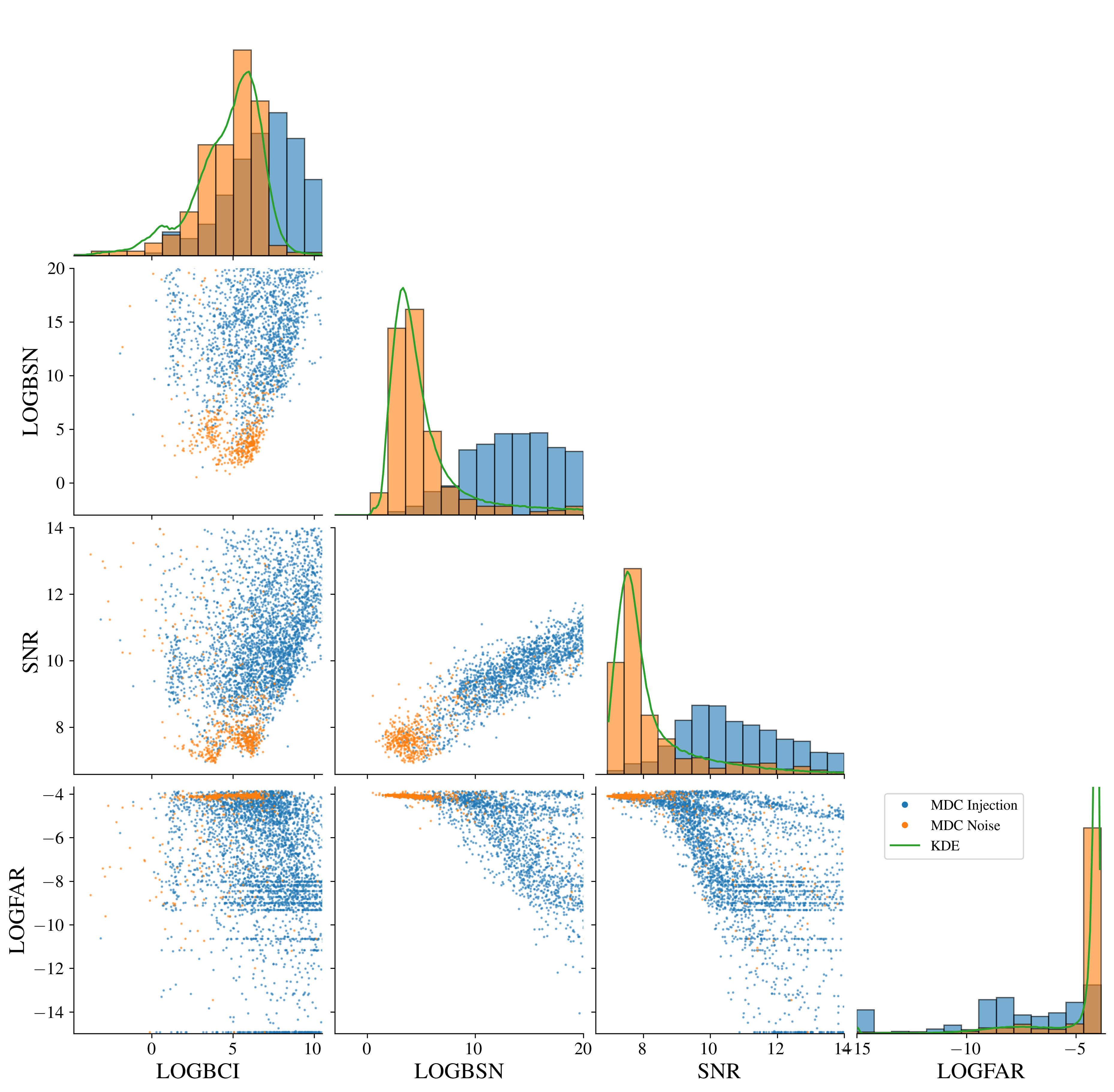}
    \caption{Corner plot of the MBTA and BAYESTAR pipeline data products from \ac{MDC} events windowed around noise events. The histograms are normalized and the green line shows the \ac{PDF} of the KDE.}
    \label{fig:MBTA_corner}
\end{figure}

\begin{figure}
    \centering
    \includegraphics[width=\linewidth]{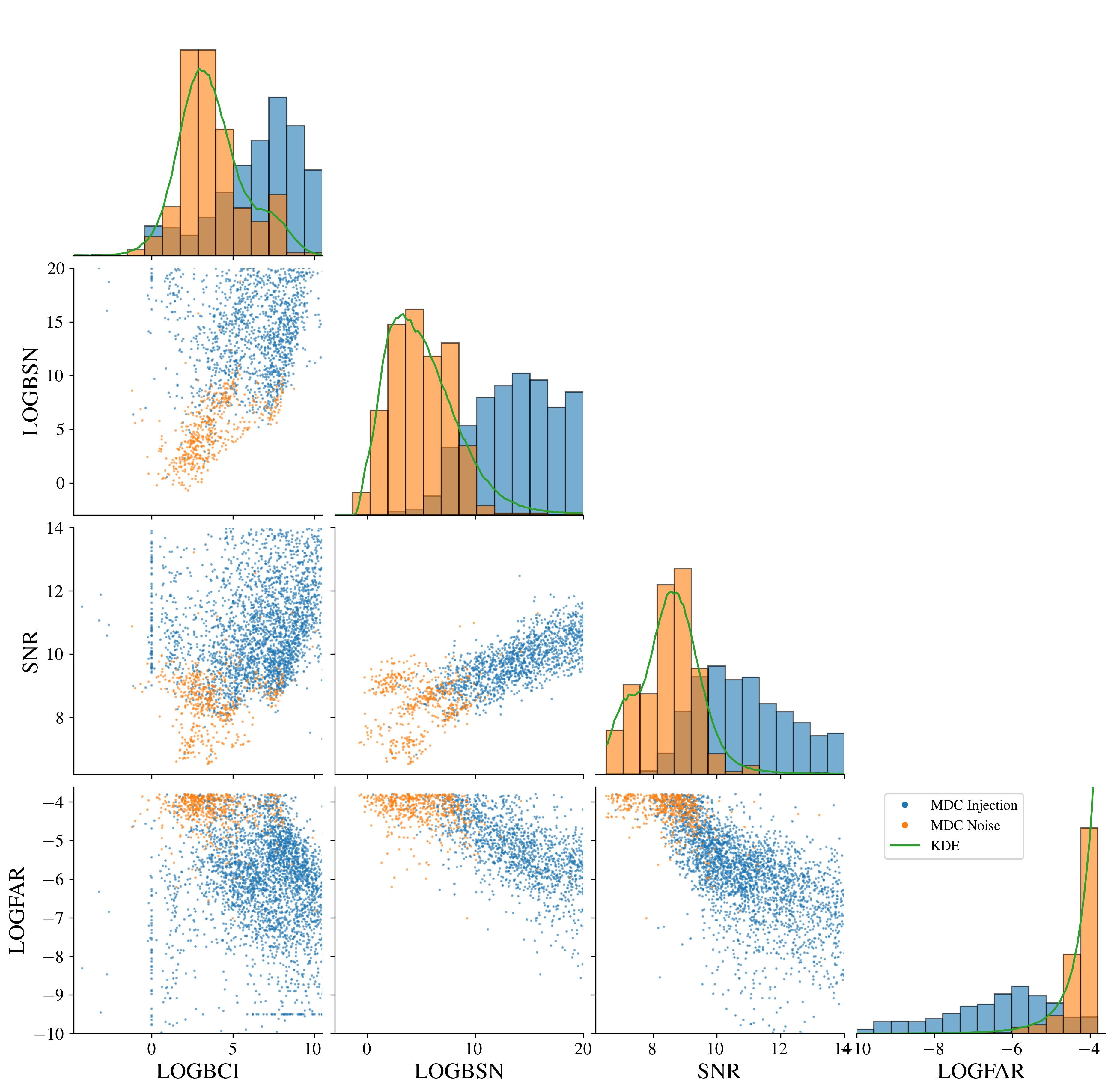}
    \caption{Corner plot of the PyCBC and BAYESTAR pipeline data products from \ac{MDC} events windowed around noise events. The histograms are normalized and the green line shows the \ac{PDF} of the KDE.}
    \label{fig:pycbc_corner}
\end{figure}

\begin{figure}
    \centering
    \includegraphics[width=\linewidth]{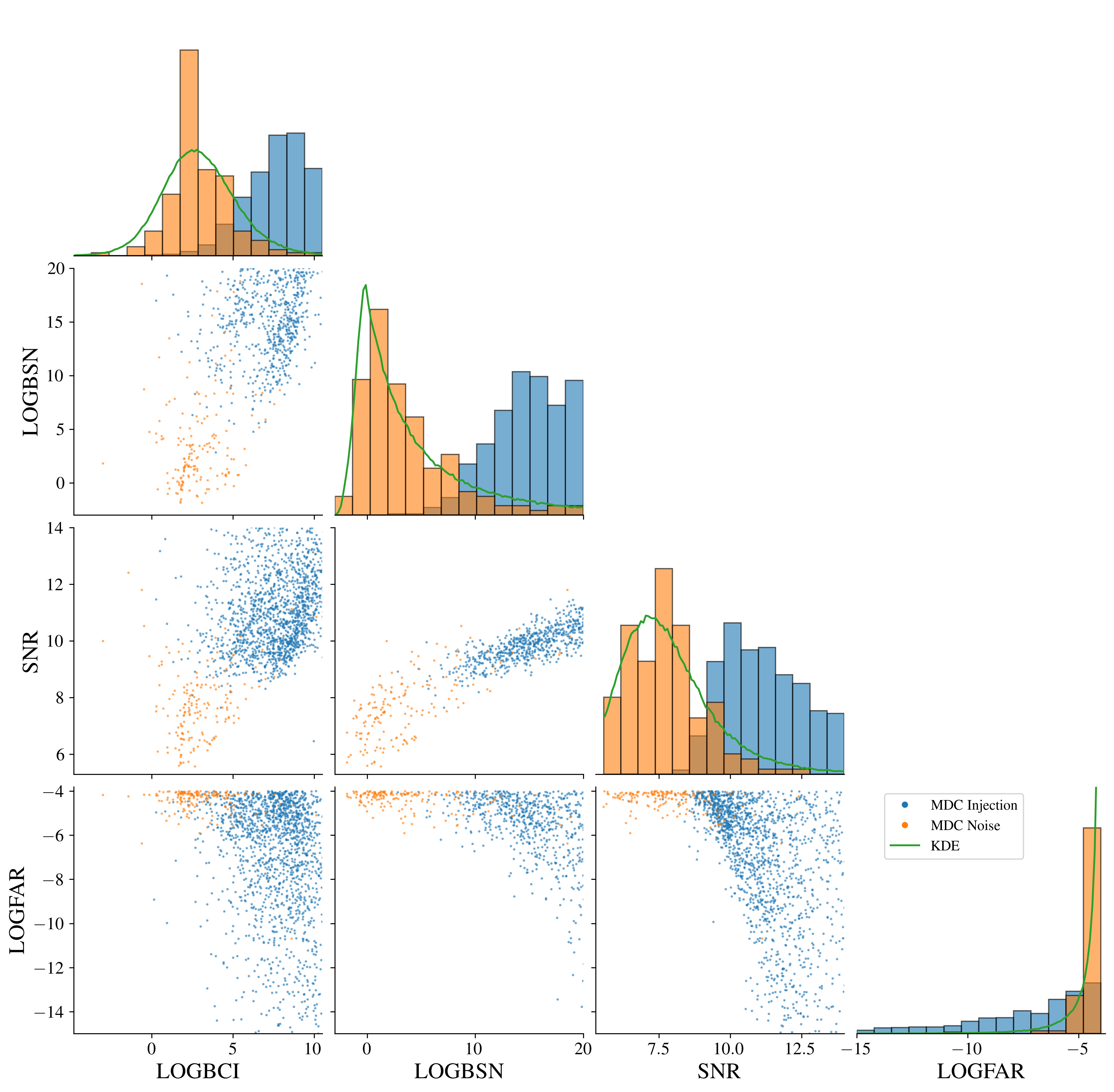}
    \caption{Corner plot of the SPIIR and \texttt{BAYESTAR} pipeline data products from \ac{MDC} events windowed around noise events. The histograms are normalized and the green line shows the \ac{PDF} of the KDE.}
    \label{fig:spiir_corner}
\end{figure}

Shown here in Figs. \ref{fig:MBTA_corner}, \ref{fig:pycbc_corner}, and \ref{fig:spiir_corner} are the corner plots corresponding to the \ac{MBTA}, \ac{PyCBC}, and \ac{SPIIR} \ac{MDC} data and \ac{KDE}s, complementary to the one shown in Fig. \ref{fig:corner_gstlal} for \ac{GstLAL}.

\end{appendix}

\end{document}